\documentclass[a4paper]{jpconf}
\usepackage{graphicx}

\newcommand{\nc}{\newcommand}           
\nc{\vc}[1]     {\mbox{\boldmath $#1$}} 
\nc{\wtil}      {\widetilde}            
\nc{\bras}[1]   {\langle#1|}            
\nc{\kets}[1]   {|#1\rangle}            
\nc{\hO}        {O}           

\begin{document}

\title{Resonances and continuum states of drip-line nuclei using the complex scaling method}

\author{Takayuki Myo$^1$ and Kiyoshi Kat\=o$^2$}

\address{$^1$General Education, Faculty of Engineering, Osaka Institute of Technology, Osaka, Osaka 535-8585, Japan}
\address{$^2$Division of Physics, Graduate School of Science, Hokkaido University, Sapporo 060-0810, Japan}

\ead{myo@ge.oit.ac.jp}

\begin{abstract}
Resonances and continuum states of He isotopes are investigated using the cluster orbital shell model (COSM) with the complex scaling method (CSM).
We discuss the following subjects:
1) Spectroscopic factors of the unbound nucleus $^7$He into the $^6$He-$n$ components
and their relation to the one-neutron removal strengths of $^7$He.
The importance of the $^6$He($2^+$) resonance is shown.
2) Structure of five-body $0^+$ resonance of $^8$He from the viewpoint of the two-neutron pair coupling.
The monopole strengths into five-body unbound states are also investigated.
It is found that the sequential breakup process of $^8$He $\to$ $^7$He+$n$ $\to$ $^6$He+$n$+$n$ is dominant in the monopole excitation,
while the contribution of $^8$He($0^+_2$) is negligible. 
\end{abstract}

\section{Introduction}

Recently, many experiments on He isotopes have been done\cite{skaza06,skaza07,golovkov09} and the exotic structures in those nuclei have been reported.
The ground states of $^6$He and $^8$He have found to exhibit the halo and skin structures of valence neutrons, 
respectively, which are loosely bound system surrounding the $\alpha$ particle. 
On the other hand, the structures of the excited states in He isotopes are not understood yet.
For He isotopes, most of the states are located above the threshold energy of the $\alpha$ emission\cite{skaza07}.
This fact indicates that the resonances of He isotopes can decay into 
various channels, such as, for $^8$He, the channels of $^7$He+$n$,  $^6$He+2$n$, $^5$He+3$n$ and $^4$He+4$n$.
These multiparticle decays make it difficult to observe the excited states of He isotopes\cite{skaza06}.

In the theory to describe the resonances, it is important to satisfy the multiparticle decay condition for all open channels.
Emphasizing this condition, we employed the cluster orbital shell model (COSM)\cite{suzuki88}
of the $^4$He+$N_{\rm v} n$ systems for He isotopes, where $N_{\rm v}$ is a valence neutron number around $^4$He.
In COSM, all open channels are taken into account explicitly, so that we can treat the many-body decaying phenomena.
In this study, we discuss the structures of the four- and five-body resonances of $^7$He and $^8$He, respectively, 
including the full couplings with $^{5,6}$He\cite{myo077,myo09,myo10}. 
The resonances are described as Gamow states using the complex scaling method (CSM)\cite{moiseyev98,aoyama06},
under the correct boundary conditions for all decay channels. 

The successful results of He isotopes are obtained for energies and decay widths.
The spectroscopic factors ($S$ factors) of $^7$He into $^6$He-$n$ are shown to clarify
the coupling between the halo state of $^6$He and a last neutron in $^7$He.
For $^8$He, we concentrate on the $0^+$ states and predict the excited five-body resonances.
We also show the monopole strength into the unbound states of $^8$He.
This quantity is useful to see the characteristics not only of the resonances, but also of non-resonant continuum states of $^8$He. 
We also clarify the contributions of two kinds of the final states of $^6$He+2$n$ and $^4$He+4$n$.
Similar analysis has been performed in the three-body Coulomb breakups of halo nuclei\cite{myo01,myo0711}.


\section{Method}
We use COSM to describe He isotopes with the following Hamiltonian\cite{myo077,myo09,myo10};
\begin{eqnarray}
	H
&=&	\sum_{i=1}^{N_{\rm v}+1}{t_i} - T_G +	\sum_{i=1}^{N_{\rm v}} V^{\alpha n}_i + \sum_{i<j}^{N_{\rm v}} V^{nn}_{ij} ,
        \label{eq:Ham}
\end{eqnarray}
where $t_i$ and $T_G$ are the kinetic energies of each particle ($n$ and $\alpha$) and of the center of mass of the total system, respectively.
The $\alpha$-$n$ interaction $V^{\alpha n}$ is given by the microscopic KKNN potential\cite{aoyama06,kanada79}.
The $n$-$n$ interaction $V^{nn}$ is given by the Minnesota potential\cite{tang78} with modification of the strength
to fit the energy of the $^6$He ground state. 

In COSM, the total wave function $\Psi^J$ with mass number $A$ and a spin $J$ is represented by the superposition of the different configurations $\Phi^J_c$ as
\begin{eqnarray}
    \Psi^J(^{A}{\rm He})
&=& \sum_c C^J_c \Phi^J_c(^{A}{\rm He}),
    \label{WF0}\qquad
    \Phi^J_c(^{A}{\rm He})
~=~ \prod_{i=1}^{N_{\rm v}} a^\dagger_{\alpha_i}|0\rangle , 
    \label{WF1}
\end{eqnarray}
where $^4$He corresponds to a vacuum $|0\rangle$.
The creation operator $a^\dagger_{\alpha}$ is for the single particle state of a valence neutron above the $^4$He core,
with the quantum number $\alpha=\{n,\ell,j\}$ in a $jj$ coupling scheme.
Here, the index $n$ represents the different radial component.
The orbit of valence neutrons is taken up to $d$-waves.
We expand the neutron radial wave function using the Gaussian expansion method\cite{aoyama06,hiyama03}.
The index $c$ represents the set of $\alpha_i$ as $c=\{\alpha_1,\cdots,\alpha_{N_{\rm v}}\}$.
We take a summation over the available configurations in Eq.~(\ref{WF0}), which have a fixed total spin $J$.
The expansion coefficients $\{C_c^J\}$ in Eq.~(\ref{WF0}) are determined 
by diagonalization of the Hamiltonian matrix elements.

The complex scaling method (CSM) is applied to describe the many-body resonances.
In CSM, we transform the relative coordinates $\vc{r}_i$ between $^4$He and $n$ in COSM, as $\vc{r}_i \to \vc{r}_i\, e^{i\theta}$
for $i=1,\cdots,N_{\rm v}$, where $\theta$ is a scaling angle.
The Hamiltonian in Eq.~(\ref{eq:Ham}) is transformed into the complex-scaled Hamiltonian $H_\theta$, and the complex-scaled Schr\"odinger equation is given as
\begin{eqnarray}
	H_\theta \Psi^J_\theta
&=&     E_\theta \Psi^J_\theta .
	\label{eq:eigen}
\end{eqnarray}
The eigenstates $\Psi^J_\theta$ are obtained by solving the eigenvalue problem of $H_\theta$ in Eq.~(\ref{eq:eigen}).
In CSM, we obtain all the energy eigenvalues $E_\theta$ of bound and unbound states on a complex energy plane, governed by the ABC theorem\cite{moiseyev98,aoyama06}.
In this theorem, it is proved that the boundary condition of Gamow resonances is transformed to the damping behavior at the asymptotic region.
This condition makes it possible to use the same method of obtaining the bound states for resonances.
The non-resonant continuum states are described being orthogonal to the Gamow resonances.
For the $^8$He case, the non-resonant continuum states of $^7$He+$n$, $^6$He+$2n$, $^5$He+$3n$ and $^4$He+$4n$ 
are obtained as well as the bound states and the resonances of $^8$He. 
These states of $^8$He consist of the extended completeness relation (ECR) using CSM\cite{myo10,aoyama06,myo01} for the five-body system as
\begin{eqnarray}
	{\bf 1}
&=&	\sum_{~\nu} \kets{\Psi_\theta^\nu}\bras{\wtil{\Psi}_\theta^\nu} , 
	\label{eq:ECR}
\end{eqnarray}
where 
$\{ \Psi^\nu_\theta,\wtil{\Psi}^\nu_\theta \}$ form a set of biorthogonal bases with a state $\nu$.
For simplicity, we here do not write the spin index explicitly. 
Similarly, for $^7$He, the four-body ECR is constructed using CSM.

We explain how to calculate the strength function using CSM.
To do this, we define the complex-scaled Green's function ${\cal G}_\theta(E)$ with the real energy $E$ of the system using ECR as
\begin{eqnarray}
	{\cal G}_\theta(E)
&=&	\frac{ {\bf 1} }{ E-H_\theta }
~=~	\sum_{~\nu}
	\frac{|\Psi_\theta^\nu\rangle \langle \wtil{\Psi}_\theta^\nu|}{E-E^\nu_\theta} , 
	\label{eq:green1}
\end{eqnarray}
where the eigenvalue $E^\nu_\theta$ is associated with the wave function $\Psi_\theta^\nu$.
The strength function ${\cal S}(E)$ from the initial state $\Psi^0$ with the operator $\hO$,
is defined in terms of Green's function as 
\begin{eqnarray}
	{\cal S}(E)
&=&     -\frac1{\pi}\ \sum_{~\nu} {\rm Im}\left[  \frac{
	\bras{\wtil{\Psi}^0_\theta}  (\hO^\dagger)_\theta \kets{\Psi^\nu_\theta}
	\bras{\wtil{\Psi}^\nu_\theta} \hO_\theta          \kets{\Psi^0_\theta}
        }{E-E^\nu_\theta}
        \right] .
	\label{eq:strength3}
\end{eqnarray}
It is noted that the function ${\cal S}(E)$ being observable, is independent of $\theta$\cite{myo09,myo01}.

\section{Results}

\subsection{Energy levels}
\begin{figure}[b]
\includegraphics[width=9.2cm,clip]{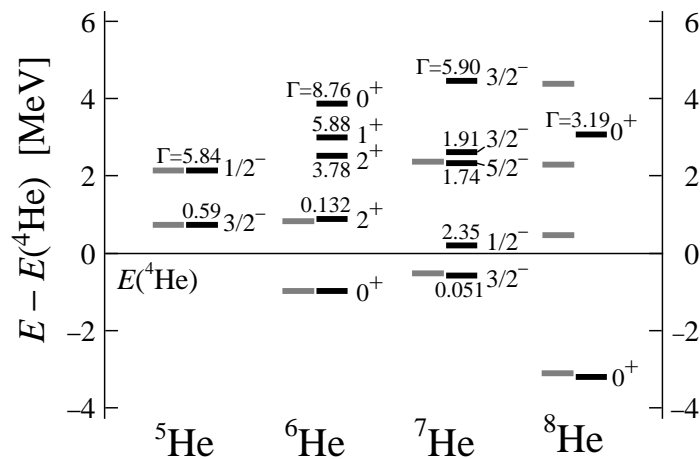}\hspace{2.0pc}%
\begin{minipage}[b]{5.8cm}
\caption{
Energy levels of He isotopes measured from the $^4$He energy. Units are in MeV.
Black and gray lines are theory and experiments, respectively. Small numbers near the black lines are theoretical total decay widths. For $^8$He, only the $0^+$ states are shown in theory.}
\label{fig:5678}
\end{minipage}
\end{figure}

We show the obtained energy spectrum of He isotopes in Fig.~\ref{fig:5678}.
One can see a good agreement between theory and experiment.
For $^7$He, the ground state is located by 0.40 MeV above the $^6$He ground state,
which agrees with the recent experiments of 0.44 MeV and 0.36 MeV\cite{skaza06}.
The $5/2^-$ state is also reproduced and other three states are predicted as four-body resonances.
For $^8$He, the ground state binding energy is obtained as 3.22 MeV from the $^4$He+$4n$ threshold, which agrees with 3.11 MeV of the experiment.
The $0^+_2$ state is predicted with the 6.29 MeV excitation energy and the 3.19 MeV decay width.

\nc{\lw}[1]{\smash{\lower1.5ex\hbox{#1}}}
\begin{table}[b]
\caption{Matter ($R_{\rm m}$) and charge ($R_{\rm ch}$) radii of $^6$He and $^8$He
in comparison with the experiments; a\cite{tanihata92}, b\cite{alkazov97}, c\cite{kiselev05}, d\cite{mueller07}.
Units are in fm.}
\label{tab:radius}
\centering
\begin{tabular}{r p{2.0cm} p{5.0cm}}
\br
                     & Present  & Experiments        \\ 
\hline
\lw{$^6$He}~~~$R_{\rm m}$  &  2.37  & 2.33(4)$^{\rm a}$~~~~2.30(7)$^{\rm b}$~~~~2.37(5)$^{\rm c}$ \\
         $R_{\rm ch}$ &  2.01  & 2.068(11)$^{\rm d}$ \\
\mr
\lw{$^8$He}~~~$R_{\rm m}$  &  2.52  & 2.49(4)$^{\rm a}$~~~~2.45(7)$^{\rm b}$~~~~2.49(4)$^{\rm c}$ \\
         $R_{\rm ch}$ &  1.92  & 1.929(26)$^{\rm d}$ \\
\br
\end{tabular}
\end{table}

The obtained matter and charge radii of $^6$He and $^8$He for their ground states reproduce the experiments,
as shown in Table~\ref{tab:radius} .
Hence, the present model well describes the neutron halo and skin structures in He isotopes.

\subsection{$^7$He}
We discuss the structures of $^7$He.
We calculate the $S$ factors of the $^6$He-$n$ components of the $^7$He resonances, listed in Table~\ref{sf}.
It is noted that the present $S$ factors correspond to the components of $^6$He inside the $^7$He resonances and contain the imaginary part\cite{myo09}.
For the $3/2^-_1$ state, the $^6$He($0^+_1$)-$n$ component almost shows a real value with a small imaginary part.
This real part well corresponds to the recent observation of $0.64\pm0.09$ \cite{beck07}.
The $^6$He($2^+_1$)-$n$ component give a large value, more than twice of that of the $^6$He($0^+_1$)-$n$ case.
For the $1/2^-$ state, the $S$ factor of $^6$He($0^+_1$)-$n(p_{1/2})$ is dominant being almost unity.
This result of $1/2^-$ indicates the weak coupling nature of the $p_{1/2}$ orbital neutron around $^6$He($0^+_1$). 
Therefore, the $1/2^-$ state of $^7$He is dominantly considered as a resonance of the $p_{1/2}$ neutron surrounding a halo state of $^6$He.

\begin{table}
\caption{$S$ factors of the $^6$He-$n$ components of $^7$He.}
\label{sf}
\begin{center}
\begin{tabular}{cll}
\br
$^7$He    &~~~~$^6$He($0^+_1$)-$n$~~~~  &~~~~$^6$He($2^+_1$)-$n$~~~~ \\ \mr
$3/2^-_1$ &~~~$0.64 +i0.06$             &~~~ $1.55-i0.31 $       \\
$3/2^-_2$ &~~~$0.005+i0.01$             &~~~ $0.95+i0.02 $       \\
$3/2^-_3$ &~~~$0.003+i0.0002$           &~~~ $0.02-i0.004$       \\
$1/2^-  $ &~~~$1.00 -i0.13$             &~~~ $0.10-i0.05 $       \\
$5/2^-  $ &~~~$0.00+i0.00$              &~~~ $0.95+i0.02 $       \\ \br
\end{tabular}
\end{center}
\end{table}

\begin{figure}
\centering
\includegraphics[width=7.5cm,clip]{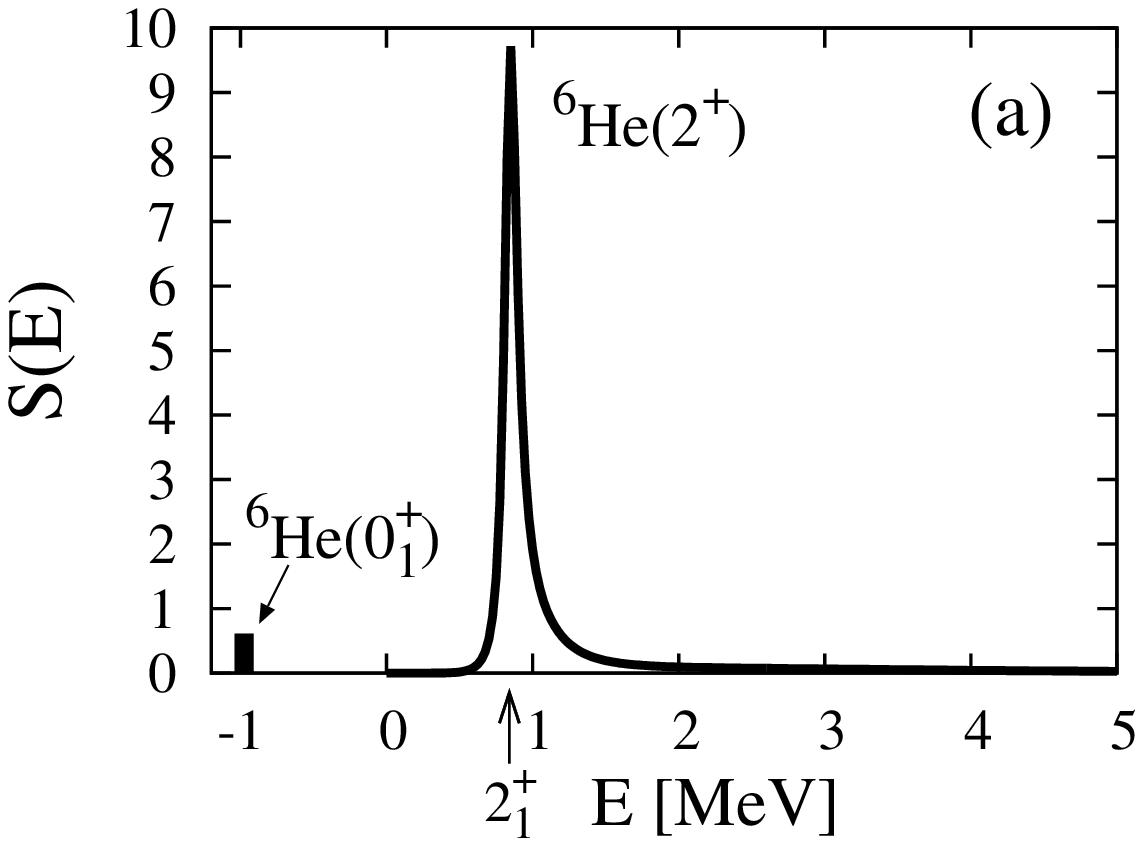}\hspace{1.0pc}%
\includegraphics[width=7.5cm,clip]{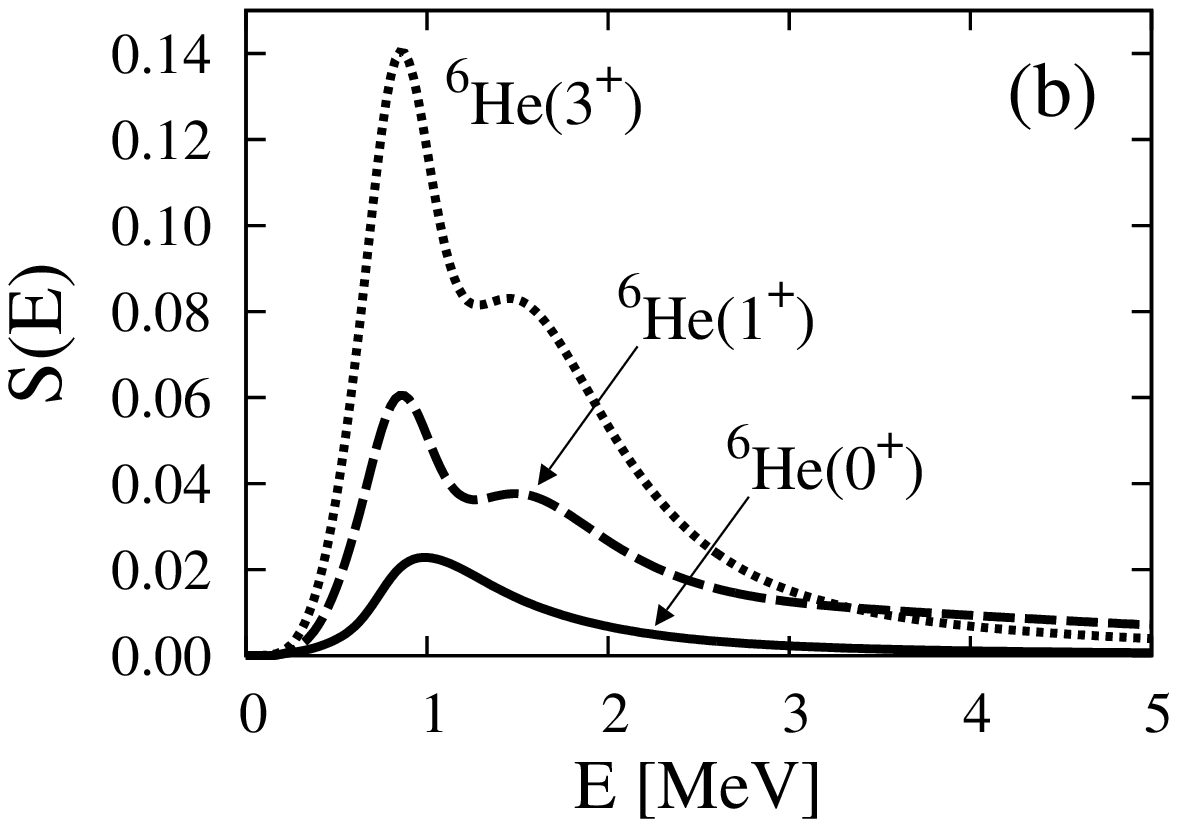}
\caption{
One-neutron removal strength of $^7$He$_{\rm g.s.}$ into $^6$He($J^\pi,E$)
states measured from the $^4$He+$n$+$n$ threshold energy. The
vertical arrow in (a) indicates the resonance energy of
$^6$He($2^+_1$). The strength into $^6$He($0^+_1$) is shown by a histogram
at the position of the corresponding energy.}
\label{fig:removal}
\end{figure}

In relation to $S$ factors, in Fig.~\ref{fig:removal}, we show the one-neutron removal strength $S(E)$ of $^7$He$(3/2^-_1)$ into $^6$He with spin $J$, 
as functions of the energy $E$ of $^6$He.
In the strength, the non-resonant continuum states of $^6$He, namely, $^5$He+$n$ and $^4$He+$n$+$n$ components
are included as well as the resonant contributions obtained in $S$ factors.
We show the results of the positive parity states of $^6$He, while the negative parity strengths give negligible contributions.
The dominant component comes from the $^6$He($2^+$) state in Fig.~\ref{fig:removal} (a), whose strength has a peak at the resonance energy of $2^+_1$.
For the energy region above the $^4$He+$n$+$n$ threshold, 
the $0^+$, $1^+$ and $3^+$ states give small contributions in the strengths, as shown in Fig.~\ref{fig:removal} (b). 
From these results, it is concluded that the one-neutron removal strength of $^7$He$(3/2^-_1)$ is dominantly exhausted 
by the $^6$He($2^+_1$) resonance above the $^4$He+$n$+$n$ threshold energy.
The residual continuum strengths of $^5$He+$n$ and $^4$He+$n$+$n$ are small.

\subsection{$^8$He}
\begin{figure}[b]
\includegraphics[width=7.7cm,clip]{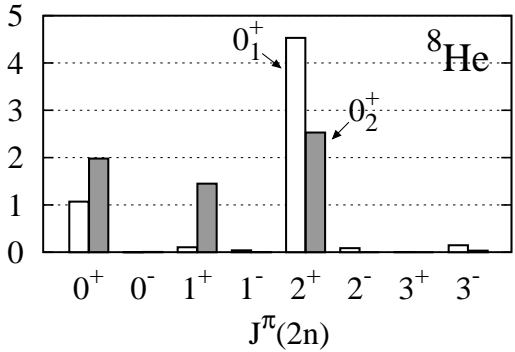}\hspace{1.8pc}%
\begin{minipage}[b]{7.0cm}
\caption{Pair numbers of the $0^+_{1,2}$ states of $^8$He.}
\label{fig:pair}
\end{minipage}
\end{figure}

We discuss the structures of the $0^+$ states of $^8$He.
We calculate the pair number of four valence neutrons in $^8$He($0^+_{1,2}$), which is defined by the matrix element
of the operator $\sum_{\alpha \le \beta} A^\dagger_{J^\pi}(\alpha\beta)A_{J^\pi}(\alpha\beta)$.
Here the quantum number $\alpha$ or $\beta$ is for the single particle state and 
$A^\dagger_{J^\pi}$ ($A_{J^\pi}$) is the creation (annihilation) operator of a single neutron-pair with spin-parity $J^\pi$.
This pair number is useful to understand the structures of four neutrons from the viewpoint of pair coupling\cite{myo10}.
In Fig.~\ref{fig:pair}, it is found that the $2^+$ neutron pair is dominant in the ground state. 
This is consistent with the configuration of $(p_{3/2})^4$ with a mixing of 86.0\%, from the CFP decomposition ( 1 and 5 for $0^+$ and $2^+$, respectively).
On the other hand, the $0^+_2$ state has much $0^+$, $1^+$ and $2^+$ states of neutron pairs.
This is also consistent with the $(p_{3/2})^2(p_{1/2})^2$ configuration with a mixing of 96.9\%,
which is decomposed into the pairs of $0^+$, $1^+$ and $2^+$ with the occupations of $2$, $1.5$ and $2.5$, respectively.
The result of the $0^+$ neutron pair in the $0^+_{1,2}$ states of $^8$He is interesting 
in relation to the dineutron-like cluster correlation\cite{enyo07}.

\begin{figure}[b]
\includegraphics[width=7.7cm,clip]{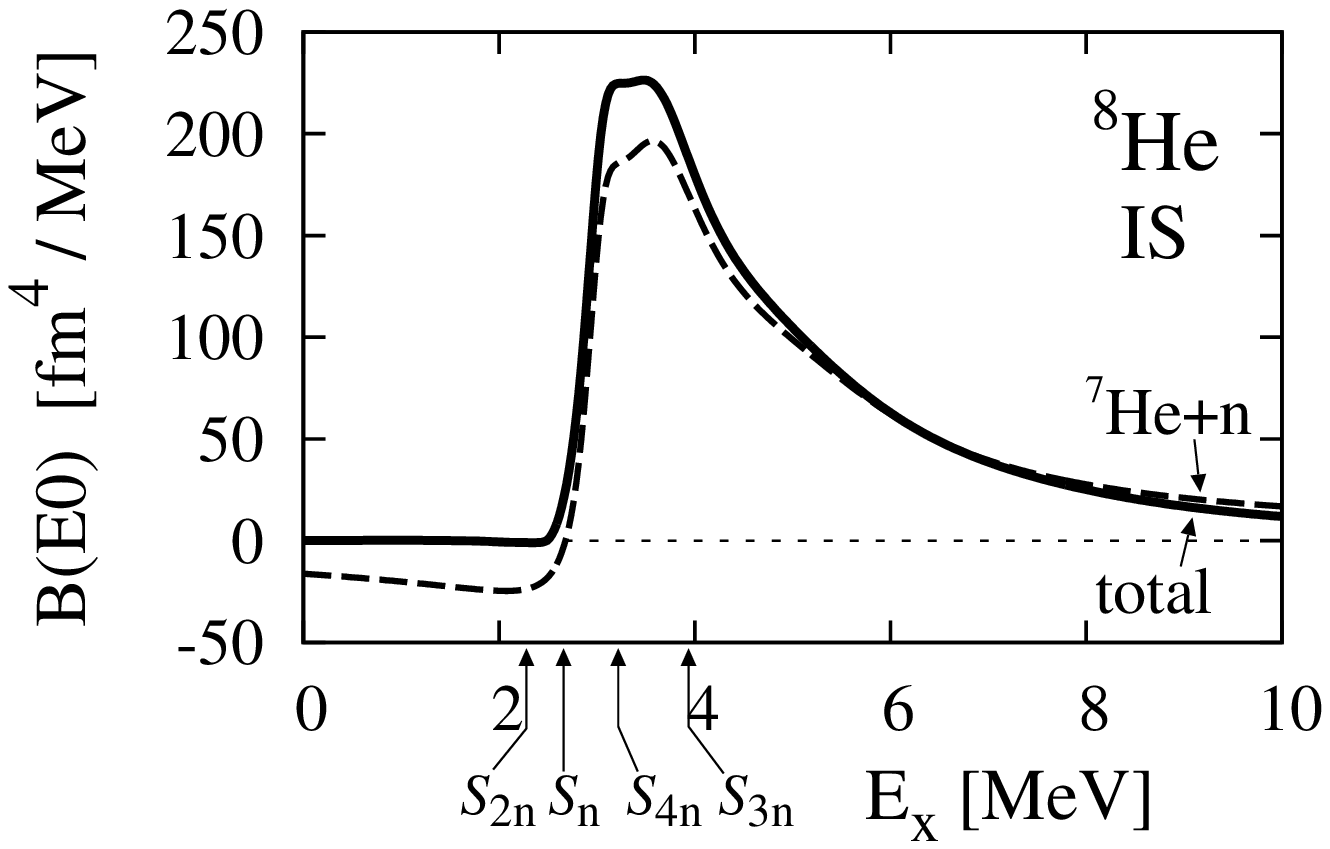}
\includegraphics[width=7.7cm,clip]{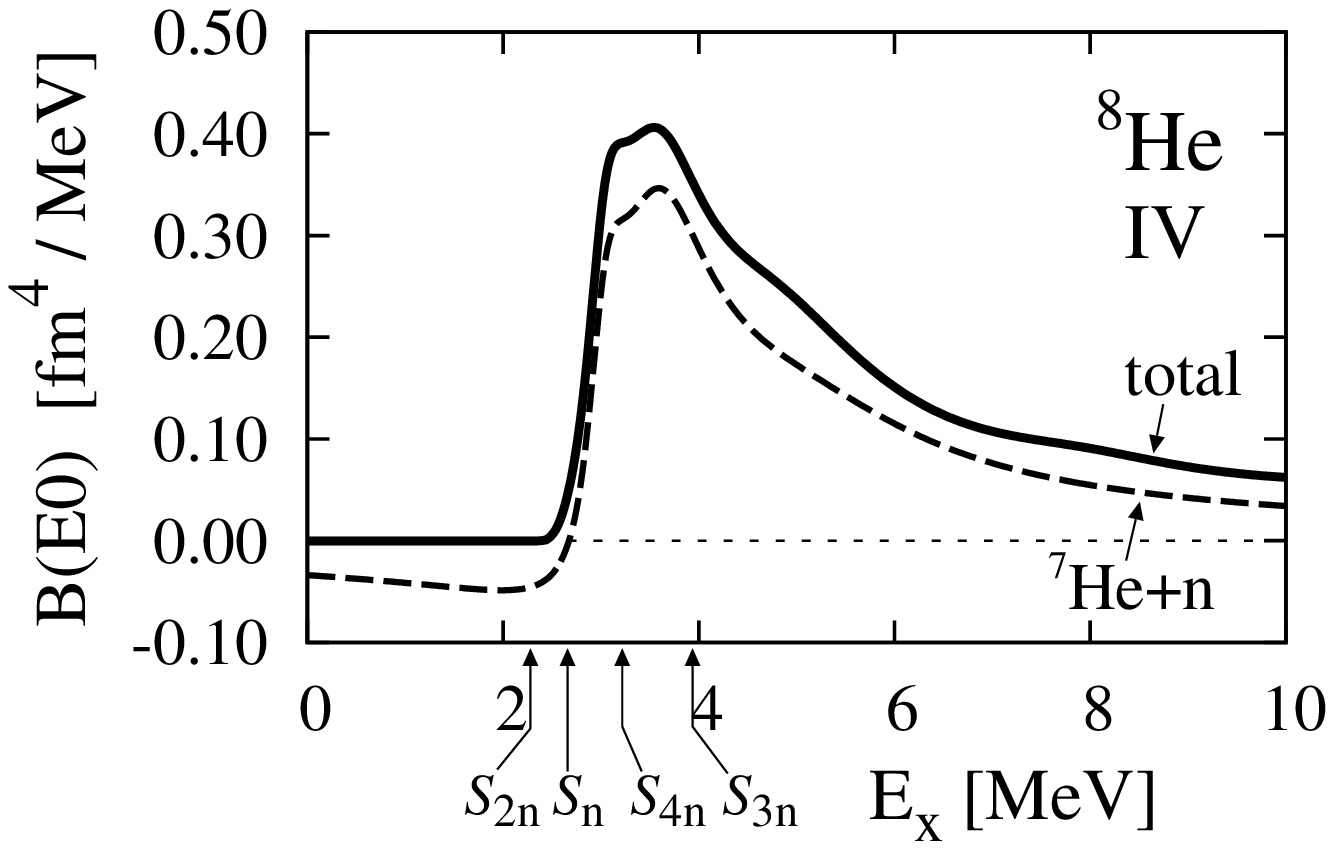}
\caption{Monopole strengths of $^8$He for isoscalar (IS) and isovector (IV) transitions as functions of the excitation energy of $^8$He.
The threshold energies of $^7$He+$n$, $^6$He+$2n$, $^5$He+$3n$ and $^4$He+$4n$ are indicated by arrows with $S_n$, $S_{2n}$, $S_{3n}$ and $S_{4n}$, respectively.}
\label{fig:monopole}
\end{figure}

Finally, we calculate the monopole transition of $^8$He and see the effect of the $0^+_2$ resonance.
In Fig.~\ref{fig:monopole}, the monopole strengths for isoscalar (IS) and isovector (IV) responses are shown.
It is found that two strengths commonly show the low energy enhancement just above the threshold energy.
There is no clear signature of the $0^+_2$ state around its excitation energy of 6.29 MeV in both strengths.
In fact, the transition matrix elements from the ground state into the $0^+_2$ Gamow state are
obtained as $1.78-i0.38$ fm$^4$ for IS and $-0.003+i0.018$ fm$^4$ for IV, respectively.
These values are so small in comparison with the total strengths.
This result is understood from the single particle structures of the $0^+_2$ state.
In the $0^+_2$ state, the $p_{1/2}$ orbit is largely mixed,
and this orbit cannot be excited from the $p_{3/2}$ orbit in the ground state by the monopole operator.
As a result, the monopole strength into $0^+_2$ becomes negligible. 
Instead, the continuum strength gives a main contribution.
To confirm the existence of $^8$He($0^+_2$), it is necessary to search for the suitable observables which are responsible for the $0^+_2$ state.
One of the candidates can be the two-neutron transfer reaction into $^6$He to produce the excited states of $^8$He.
Experimentally, the $^6$He($t$,$p$)$^8$He reaction was reported without spin assignment\cite{golovkov09}.

We extract the continuum components of the monopole strengths in Fig.~\ref{fig:monopole}.
It is found that the two strengths dominantly come from the $^7$He($3/2^-_1$)+$n$ components.
This selectivity of the continuum states is related to the properties of the monopole operator, which excites one of the relative motions in $^8$He.
As a result, the intercluster motion between the $^7$He cluster and a valence neutron is strongly excited from the $^8$He ground state.
This result indicates the sequential breakup process of $^8$He(G.S.) $\to$ $^7$He($3/2^-_1$)+$n$ $\to$ $^6$He(G.S.)+$n$+$n$ in the monopole excitations,
instead of the $0^+_2$ resonance and the $^4$He+$4n$ five-body breakup states.

\section{Summary}

We have investigated the structures of $^{7,8}$He using the five-body cluster model with the correct boundary condition for resonances.
The calculated $S$ factors of $^7$He indicate that the $1/2^-$ resonance is dominantly the weakly coupled state consisting of 
a halo state of $^6$He and a valence neutron.
The one-neutron removal strength of $^7$He is successfully obtained and the importance of the $^6$He($2^+_1$)+$n$ channel is shown.
We also predict the five-body $0^+_2$ resonance of $^8$He, which dominantly has a $(p_{3/2})^2(p_{1/2})^2$ configuration.
We further investigate the monopole strengths of $^8$He in the breakup energy region.
It is dominant that the sequential breakup process of $^8$He via the $^7$He+$n$ states into the $^6$He+$2n$ three-body final states.

\ack
This work was supported by a Grant-in-Aid for Young Scientists from the Japan Society for the Promotion of Science 
(No. 21740194).
Numerical calculations were performed on a supercomputer (NEC SX8R) at RCNP, Osaka University.


\end{document}